# Thermodynamic Origins of Structural Metastability in Two-Dimensional Black Arsenic

*Guoshuai Du,*[1,2] *Feng Ke,*[3,4] *Wuxiao Han,*[1,2] *Bin Chen,*[4] *Qinglin Xia,*[5,*] *Jun Kang,*[6,*]

*and Yabin Chen*[1,2,7,*]

[1]School of Aerospace Engineering, Beijing Institute of Technology, Beijing 100081, China

[2]Advanced Research Institute of Multidisciplinary Science, Beijing Institute of Technology, Beijing 100081, China

[3]Department of Geological Sciences, Stanford University, Stanford, CA 94305, USA

[4]Center for High Pressure Science and Technology Advanced Research, Shanghai 201203, China

[5]School of Physics and Electronics, Hunan Key Laboratory of Nanophotonics and Devices, Central South University, Changsha, 410083 China

[6]Beijing Computational Science Research Center, Beijing 100193, China

[7]BIT Chongqing Institute of Microelectronics and Microsystems, Chongqing, 400030, China

[*]E-mail: qlxia@csu.edu.cn, jkang@csrc.ac.cn, chyb0422@bit.edu.cn

**ABSTRACT:** Two-dimensional (2D) materials have aroused considerable research interests owing to their potential applications in nanoelectronics and optoelectronics. Thermodynamic stability of 2D structures inevitably affects the performance and power consumption of the fabricated nanodevices. Black arsenic (b-As), as a cousin of black phosphorus, has presented the extremely high anisotropy in physical properties. However, the systematic research on structural stability of b-As is still lack. Herein, we demonstrated the detailed analysis on structural metastability of the natural b-As, and determined its existence conditions in terms of two essential thermodynamic variables as hydrostatic pressure and temperature. Our results confirmed that b-As can only survive below 0.7 GPa, and then irreversibly transform to gray arsenic, in consistent with our theoretical calculations. Furthermore, thermal annealing strategy was developed to precisely control the thickness of b-As flake, and it sublimates at 300 °C. These results could pave the way for 2D b-As in many promising applications.

**KEYWORDS:** black arsenic, metastable, thermodynamics, diamond anvil cell, phase transition, thermal etching

**Table of Contents:**

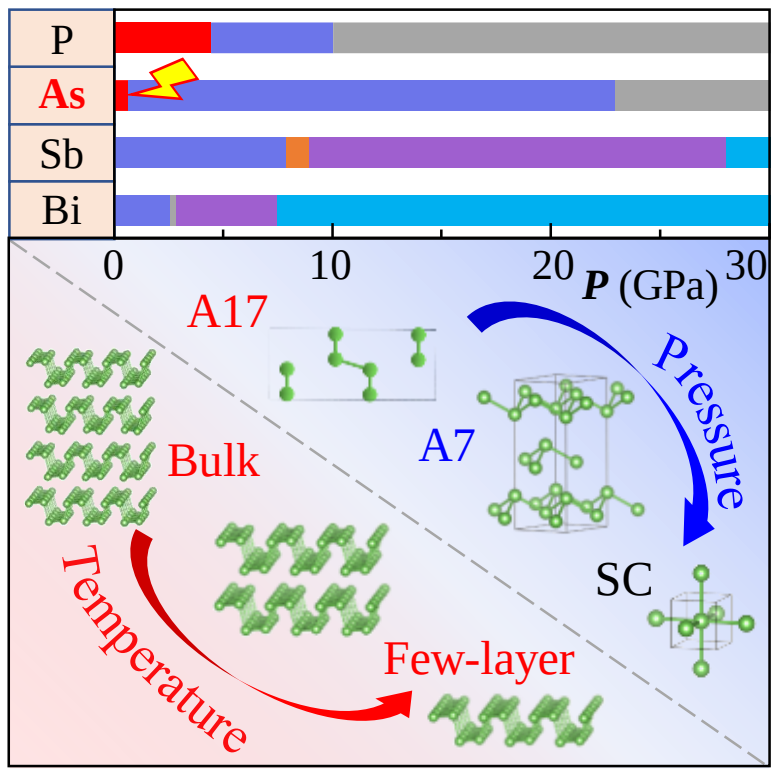

Two-dimensional (2D) layered nanomaterials emerge and have attracted extensive attention due to their exotic physical and chemical properties,[1-10] since the successful mechanical exfoliation of graphene.[11] Ultrathin 2D semiconductors, such as transition metal dichalcogenides,[12] black phosphorus (b-P)[13] and MXenes,[14] are regarded as the competitive candidate materials for the next-generation nanoelectronics because of their atomic thickness, unique flexibility, and tunable band structures.[15, 16] For the practical applications, it is well known that the structural stability of 2D materials plays a critical role and determines the process yield, reliability and lifetime of the fabricated nanodevices.[17, 18] Based on the atomic thickness of 2D materials, the various external stimuli can sensitively affect their phase structures and further seriously impact the device performances, including temperature, stress, optical illumination, and humidity.[17] For instance, b-P, supposed for mid-infrared photodetectors[19, 20] and field-effect transistors,[21] severely suffers from the chemical degradation, photooxidation[22] and surface instability in ambient conditions.[23] Although several passivation methods have developed,[24, 26] the encapsulated or modified b-P flakes by other coatings generally exhibit the limited detectivity or carrier mobility. Many 2D magnets[27] and the relative devices, e.g. few-layer $CrI_3$[28] and $VSe_2$[29] cannot survive even in air.

Black arsenic (b-As), as a cousin of b-P, is reported to be a novel 2D layered semiconductor with the extremely anisotropic properties.[30] With the puckered structure, we found that b-As can present the orientation-dependent Raman spectrum, thermal conductivity and carrier transport behaviors.[30, 31] Theoretical simulation also predicted the auxetic effect of b-As flakes, suggesting the anomalous negative Poisson's ratio.[32, 33] Despite much effort, the stability and structural transformation of b-As still remains controversial. As reported, the field-effect transistor based on few-layer b-As can function well even after exposing in air for one month.[34] While, it is also found that b-As showed high sensitivity to moisture and degraded rapidly at ambient conditions.[30] Therefore, it is of significant importance to investigate the thermodynamic stability and thus elucidate the mechanics of b-As for the prospective applications.

In thermodynamics, pressure and temperature are two fundamental variables that can influence the material stability and further induce structural phase transition. In principle, the external pressure can effectively input energy to a system by reducing its volume. Hydrostatic pressure applied by diamond anvil cell (DAC) are widely employed to modulate the lattice structure, metal-insulator transition, and high-temperature superconductors. For pnictogens of P, As, Sb and Bi in group VA, many studies have demonstrated the plenty of pressure-induced phase transitions and novel properties,[35, 36] as illustrated in Figure 1. Except b-As, the phase transitions of b-P, antimony and bismuth are irreversible as reported in Figure 1b, that is, the initial crystal structure comes back once the pressure is totally released. B-P, the steady allotrope with minimum thermodynamic energy, can undergo two obvious phase transitions from orthorhombic (O) *A*17 phase to buckled *A*7 at 4.5 GPa, further to simple cubic (SC) at 10.0 GPa.[37] For Sb and Bi, their stable phase under ambient condition is rhombohedral (R). The initial Bi can successively transform to monoclinic (M), tetragonal (Te), and body-centered cubic (BCC) phases at 8.0, 9.0 and 28.0 GPa, respectively. In contrast, the critical pressures of the heavier Bi are relatively lower. Most researches on As are focused on gray arsenic (g-As) to SC transition, occurred at 22.0 GPa.[36, 38] In addition, Gao et al. studied the transformation of b-As to g-As under high pressure.[39] However, due to the defects in the samples or the non-hydrostatic pressure conditions, the coexistence of *A*17 and *A*7 phases exists in phase transformation process. Taking b-As as the initial research object, systematic research from *A*17 to SC phase needs to be explored.

Besides pressure, thermal effect on the matter stability and device performance is of significant as well. High temperature can dramatically enhance the internal energy, and also promote the reaction rate of a system. In this regard, many 2D materials can be remarkably thinned down to exactly control the thickness at high temperature. As reported, graphene can react with oxygen to form on-site $sp^3$ defects at about 520 °C, however, bilayer graphene and thicker graphite remain intact at 600 °C.[40] In vacuum, decomposition of b-P occurred at ~400 °C,[41] so the inert $Al_2O_3$[42] or $HfO_2$[43] coatings were developed to improve the thermal stability of b-P. The calculated thermodynamic energy of b-As is higher than that of g-As by +2.5 kJ/mol,[44] indicating the metastable characteristic of b-As. As far as we know, there is still lack of systematic research on the thermal stability of the intrinsic b-As.

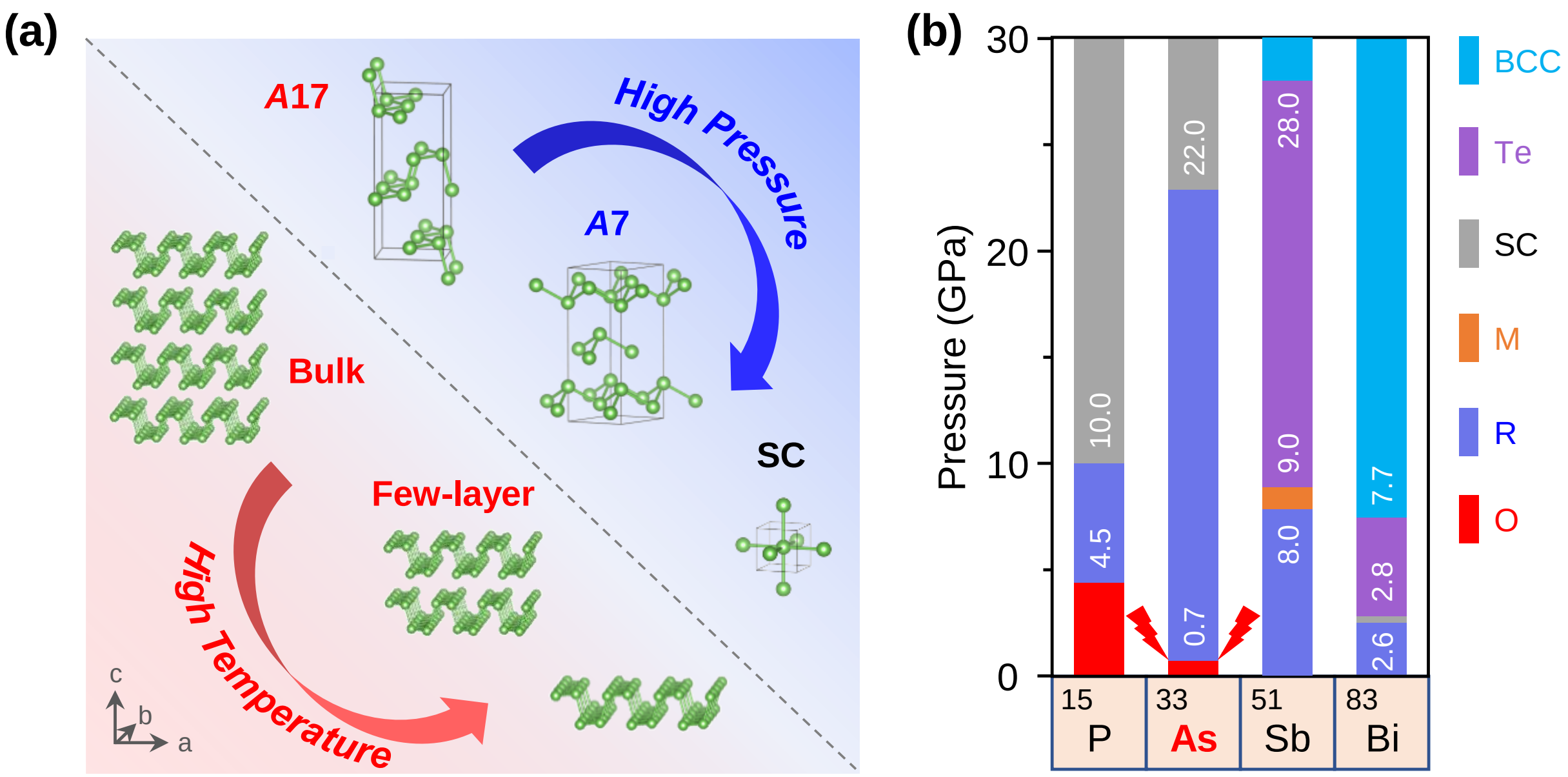

**Figure 1.** Thermodynamic stability of 2D b-As under high pressure and temperature conditions. (a) Schematic illustration of the hydrostatic pressure-induced phase transition and thermal etching of b-As flakes. (b) Pressure-induced phase transition of pnictogens in group-VA within 30 GPa. The critical pressures are well labeled. BCC, Te, SC, M, R and O mean the body-centered cubic, tetragonal, simple cubic, monoclinic, rhombohedral and orthorhombic structures, respectively.

In this work, we have confirmed the structural metastability of the natural b-As flakes, and further probed its thermodynamic origins, taking the thermodynamic variables of pressure and temperature into consideration. Under hydrostatic pressure applied by DAC, we observed two obvious phase transformations from the semiconducting b-As *A*17 (orthorhombic) to the metallic g-As *A*7 (rhombohedral) irreversibly at 0.7 GPa, and then to SC phase at 21.6 GPa. Our experimental and theoretical results confirmed that g-As structure, rather than the original b-As phase, remained when pressure was completely released, indicating the metastable nature of b-As. More uniquely, the first transition pressure around 0.7 GPa of b-As is remarkably lower that other elements in group VA. As to temperature effect, we found that the intrinsic b-As can survive at ~300 °C, and the thermal etching method was rationally developed to exactly control the thickness of a given b-As flake. We believe that these results could shed much light on the thermodynamic stability and structural transition of 2D layered materials.

In general, elemental arsenic can exist in three typical allotropes, which are g-As (*A*7 phase), yellow arsenic (y-As), and b-As (*A*17 phase).[45] In thermal equilibrium the most stable g-As phase[46] (rhombohedral, space group $R\bar{3}m$) exhibits a similar buckled structure as the layered blue phosphorus, and its indirect band gap ~1.93 eV in monolayer limit quickly closes up for trilayer or thicker g-As.[47, 48] In contrast, y-As (tetrahedral, $As_4$) adopts an insulating molecular structure alike to white phosphorus, and it is toxic and gradually transforms to g-As at ambient conditions.[49] Interestingly, b-As (orthorhombic, space group *Cmce*) forms a puckered layered structure analogous to b-P, and its moderate band gap decreases from 0.83 eV in monolayer to 0.31 eV in bulk.[30, 50]

To explore pressure-effect on the stability of b-As, we performed in situ Raman scattering and X-ray diffraction (XRD) measurements of b-As under high pressure applied by DAC. We first use Raman spectroscopy to efficiently distinguish the various phases and structures of arsenic allotropes, especially b-As and g-As. According to symmetry analysis, the point group of bulk b-As is $D_{2h}$, and four atoms in each primitive cell correspond to twelve normal vibrational modes at Γ point in Brillouin zone, including three acoustic branches and nine optical branches.[51] The irreducible representation of lattice vibrational modes for bulk b-As can be represented as $\Gamma = 2A_g \oplus B_{1g} \oplus B_{2g} \oplus 2B_{3g} \oplus A_u \oplus 2B_{1u} \oplus 2B_{2u} \oplus B_{3u}$, $2A_g$, $B_{1g}$, $B_{2g}$ and $2B_{3g}$ modes of which are Raman active. When following backscattering configuration and the excitation laser polarized along $c$ axis of b-As specimen, $A_g^1$, $B_{2g}$ and $A_g^2$ modes along out-of-plane, zigzag, and armchair orientation could be detected, respectively. As shown in Figure 2a and 2b, before any high-pressure treatment of $A17$ phase (b-As), their Raman shifts observed at 219.8, 226.5, and 253.9 cm$^{-1}$ are well consistent with the literature.[39] In comparison, the $A7$ phase of g-As belongs to the $D_{3d}$ point group,[52] leading to three Raman active modes. In Figure 2c, the two-fold degenerate $E_g$ and $A_{1g}$ vibrations are parallel and perpendicular to the crystal layer, respectively. In our case, the $A7$ sample was acquired through the irreversible phase transition of b-As under high pressure as discussed below. The extracted Raman shifts at 196.2 and 255.8 cm$^{-1}$ are fairly attributed to $E_g$ and $A_{1g}$ modes, respectively. Due to the greater relaxation rates of $E_g$ mode, its full width at half-maximum is obviously boarder than $A_{1g}$ mode, which is observed in natural g-As as well.[52, 53]

In situ Raman characterizations were performed to explore the structural transitions of b-As under hydrostatic pressure. Figure 2d-e present the pressure-dependent Raman spectra of b-As during compression and the sequent decompression processes. As the pressure gradually increased up to 29.4 GPa, b-As underwent two distinct structural phase transitions from $A17$ to $A7$ at 0.7

GPa, and further to simple cubic (SC) phase around 22 GPa. When the pressure was slowly released to ambient pressure, the $A7$ phase with lower free energy was preserved other than the original $A17$ phase. The irreversible phase transition undoubtedly validates the thermodynamic metastability of $A17$ phase as predicted previously.

In details, at pressure lower than 0.8 GPa, three Raman peaks $A_g^1$, $B_{2g}$ and $A_g^2$ of b-As displayed a clear blueshift with increasing of pressure as shown in Figure 2d and the inset of Figure 2e. Moreover, the out-of-plane mode $A_g^1$ has the largest linear slope, $\partial\omega/\partial P$ as 6.17 cm$^{-1}$/GPa, which is attributed to the higher compressibility along out-of-plane orientation than that along in-plane orientation. The out-of-plane interaction is governed by van der Waals forces, much weaker than As-As bonds in the crystalline plane. The $\partial\omega/\partial P$ slopes of $B_{2g}$ and $A_g^2$ are fitted as 3.30 and 0.08 cm$^{-1}$/GPa, respectively. Compared with the larger $\partial\omega/\partial P$ of $B_{2g}$, the frequency of $A_g^2$ mode remains nearly constant when pressure lower than 0.8 GPa, indicating the small lattice expansion along the armchair direction.

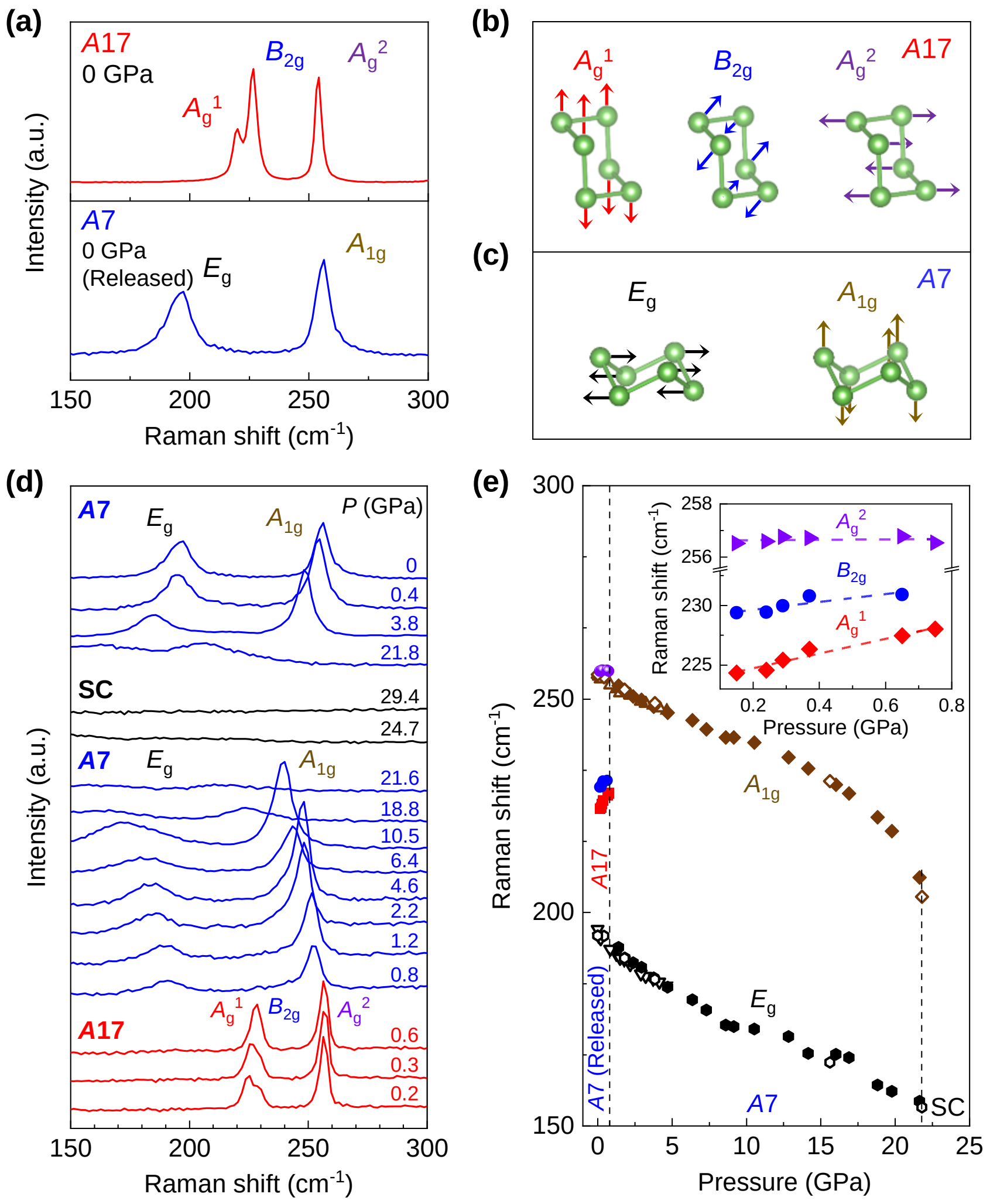

**Figure 2.** Hydrostatic pressure effect on the structural stability of b-As and the relative Raman measurements. (a) Raman spectra of $A17$ phase (upper) and $A7$ phase (lower) at ambient pressure. (b) Atomic displacement schemes of Raman vibration modes for $A17$ phase. $A_g^1$, $B_{2g}$ and $A_g^2$ vibrate along out-of-plane, zigzag, and armchair directions, respectively. (c) Atomic displacement schemes of Raman vibration modes for $A7$ phase. $E_g$ and $A_{1g}$ modes vibrate along in-plane and out-of-plane directions, respectively. (d) Pressure-dependent Raman spectra of b-As up to 29.4 GPa. Two structural transitions happen from $A17$ to $A7$ at 0.7 GPa, and further to SC at 22 GPa. (e) Pressure dependence of phonon frequencies of $A17$ and $A7$ phases. Inset is the detailed evolution of Raman shift of b-As with pressure lower than 0.8 GPa.

At pressure between 0.8 and 21.6 GPa, we found that $A17$ phase was fully transformed to $A7$ phase. When pressure increased to 0.8 GPa, three Raman modes $A_g^1$ , $B_{2g}$ and $A_g^2$ of b-As suddenly vanished, and two new Raman modes appeared at 190.2 $cm^{-1}$ for $E_g$ and 252.3 $cm^{-1}$ for $A_{1g}$ mode of g-As, as shown in Figure 2d. These two new modes of g-As became redshift and broadened together with increased pressure, in agreement with our density functional theory (DFT) calculation in Figure S1. This result further confirmed that $A7$ to SC transition in As could be considered as a prototype example for the pressure-induced suppression of Peierls-type distortion in layered elemental state, involving a semimetal-to-metal transition as well as the strong phonon coupling and phonon softening.[36, 38] When pressure exceeds 21.6 GPa, all Raman signals disappeared completely, suggesting a new metallic phase with the abundant carrier concentration. In view of this, the first-order phase transition from rhombohedral $A7$ to SC phase is expected.[38] During the decompression process, $E_g$ and $A_{1g}$ modes of g-As started to emerge at 21.8 GPa, and remained even back to ambient condition.

To determine the crystal structure and evaluate lattice parameter of b-As, we deployed the synchrotron XRD measurements of b-As under high pressure at ambient temperature. The maximum pressure was up to 38.9 GPa. As shown in Figure 3, the evolution of XRD data presents the discontinuous changes at approximately 0.4 and 25.0 GPa, revealing the occurrence of phase transitions. Accordingly, b-As first transformed completely to g-As at ~0.4 GPa and then to SC phase at ~25.0 GPa during compression. On decompression, the sample remained in g-As until the pressure was totally released to ambient pressure, well consistent with Raman measurements in Figure 2. All above results are strong evidences on the thermodynamic metastability of b-As.

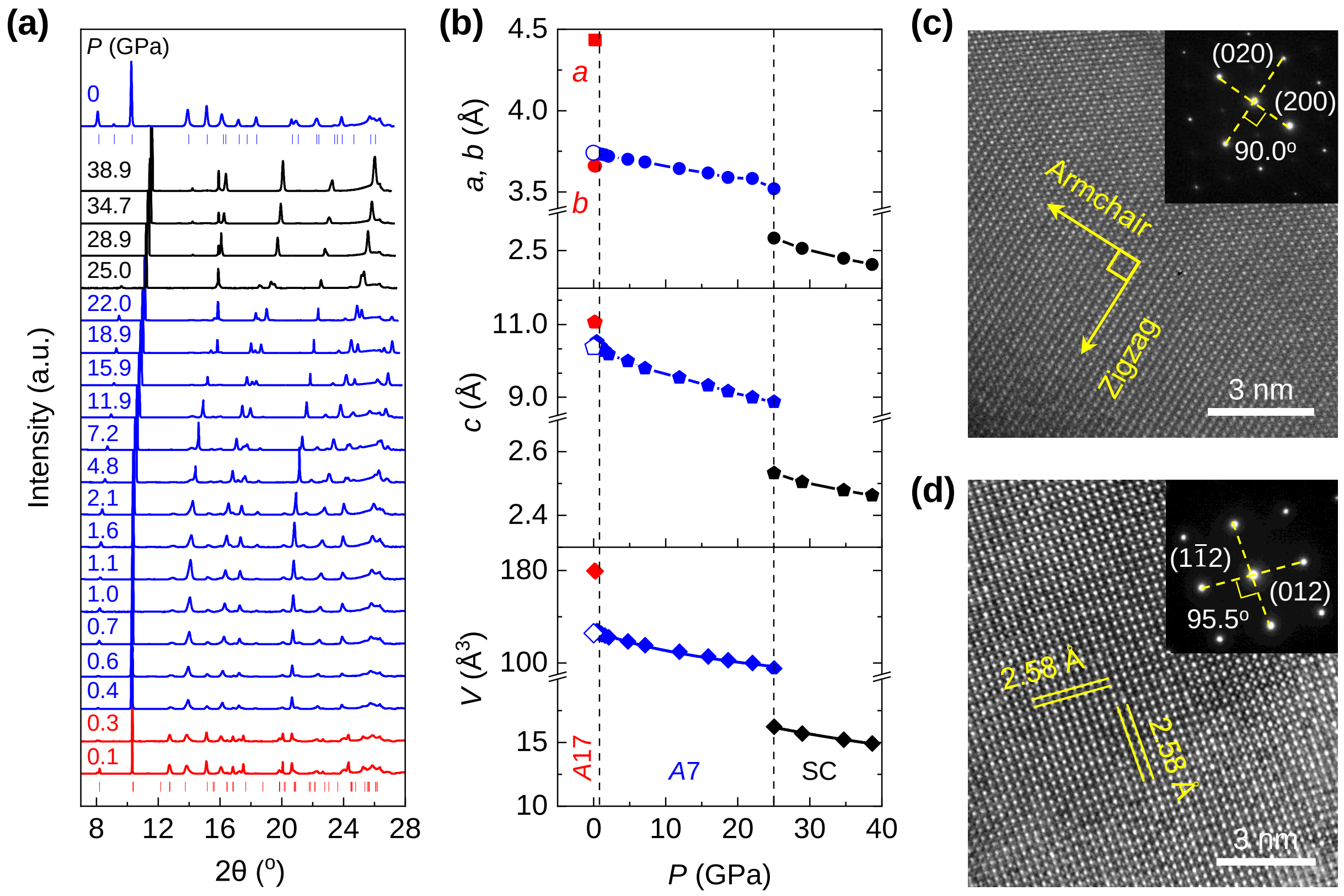

**Figure 3.** Pressure-induced structural phase transition of b-As and the XRD and TEM characterizations under high pressure. (a) XRD curves of b-As during compression and decompression processes. (b) Pressure dependence of the lattice parameters and volume V of b-As. The solids and hollow points are obtained from compression and decompression process, respectively. The dashed lines represent the third-order fitted Birch-Murnaghan EOS fitting to the measured P-V data. (c) High-resolution TEM image of b-As before the compression. Inset is the indexed pattern of selected area electron diffraction (SAED). (d) High-resolution TEM image of g-As, which is taken out from DAC chamber after the compression. Inset is the representative SAED pattern.

We further determined pressure-dependent lattice parameters and cell volumes ($V$) for the $A17$, $A7$ and SC structures. For orthorhombic $A17$ phase, $a \neq b \neq c$. By contrast, for rhombohedral $A7$ phase, $a = b \neq c$. It is noted that *a* and *b* are the in-plane lattice parameters, *c* is the out-of-plane lattice parameter. As shown in Figure 3b, the change trend of lattice parameters with pressure obviously proves the occurrence of phase transitions, and all lattice parameters rationally decreased with pressure, accompanying that the volume significantly contracts. For SC phase ($a = b= c$), the slight decrease of lattice parameters benefits from its difficult compressibility under high pressure. The distinct discontinuities of lattice parameters and $V$ at the critical pressure evidently interpret the first-order phase transitions. Furthermore, we obtained the bulk modulus $B_0$ of *A*7 and SC phases by fitting the pressure-volume data with the third-order Birch-Murnaghan equation of state (EOS): $P(V) = \frac{3B_0}{2}\left[\left(\frac{V_0}{V}\right)^{\frac{7}{3}} - \left(\frac{V_0}{V}\right)^{\frac{5}{3}}\right]\left\{1 + \frac{3}{4}\left(B_0^{'} - 4\right)\left[\left(\frac{V_0}{V}\right)^{\frac{2}{3}} - 1\right]\right\}$, where $V_0$ is the zero-pressure volume, $B_0$ is the bulk modulus at ambient pressure, and $B_0^{'}$ is a parameter for pressure derivative. With $B_0^{'}$ fixed as 4, the isothermal bulk modulus $B_0$ for *A*7 and SC phases nanosheets was extracted as 50.1±3 and 54.3±8 GPa, which is reasonably less than that of b-P (73.4±5 GPa for *A*7 phase; 147.0±2 GPa for SC phase).[54]

In addition, high-resolution transmission electron microscopy (TEM) was utilized to characterize the detailed atomic structure of the specimens before and after compression process. As described previously, *A*17 phase of arsenic is located at a local minima of the energy landscapes.[44] Thus, it is difficult to obtain pure b-As by artificial synthesis. The b-As used in this work is from nature minerals. Shown in Figure 3c is the atomic-resolution TEM image, where two orthogonal arrows denote the zigzag and armchair directions. The extracted lattice parameter, $a$ = 4.6 Å and $b$ = 3.5 Å are consistent with XRD results. After the compression, the sample taken out of DAC chamber also presented crystalline structure, and the interplanar crystal spacing approximates 2.58 Å, shown in Figure 3d. Importantly, the zone axis perpendicular to the sample surface changed from [001] of $A17$ to $[\bar{4}\bar{2}1]$ of $A7$. The crystal surface changes from $A17$ to $A7$ in SAED are definitely identical with our calculated results by DFT, as shown in Figure S1. It is noticed that we also found the two-phase coexistence ($A17$ and $A7$) at some samples, as shown in Figure S6, owing to the inevitable defects or non-uniform hydrostatic condition.[55]

To further clarify the phase transition mechanism of b-As under high pressure, theoretical calculations were performed based on DFT. As shown in Figure S1, phase transition from the orthorhombic $A17$ to rhombohedral $A7$ phase is attributed to the intralayer bond breakage and interlayer bond formation.[39] Thus, we could predict that the phase transition from $A17$ to $A7$ could only occur in bulk and few-layer b-As, just like the scenario as b-P.[54] For g-As, two Raman vibration modes ($E_g$ and $A_{1g}$) decrease in frequency, and the rhombohedral angle $\alpha$ increases gradually to ~60° under compression, which is exactly the cubic structure.[38] Besides, the fractional coordinate $z$ also increases throughout the pressure range of $A7$ structure in Figure S1c. With the decrease of $c/a$ ration of $A7$ structure, the interlayer and intralayer distances become less different and finally the SC structure is approached.[35] Moreover, rhombohedral $A7$ structure can be derived from SC structure by a rhombohedral distortion along the [111] direction, and a simultaneously relative displacement of (111) planes towards each other in pairs along [111] direction, and could be represented by four distorted SC lattices.[38, 56] Also, each arsenic atom has three short contacts in the layer and three longer contacts with the As atoms of the adjacent layer.[35] Under compression, the Peierls-type pairing distortion of the $A7$ structure decreases and ultimately vanishes, that is the SC lattices is no longer distorted, resulting in the phase transition from $A7$ to SC structure.

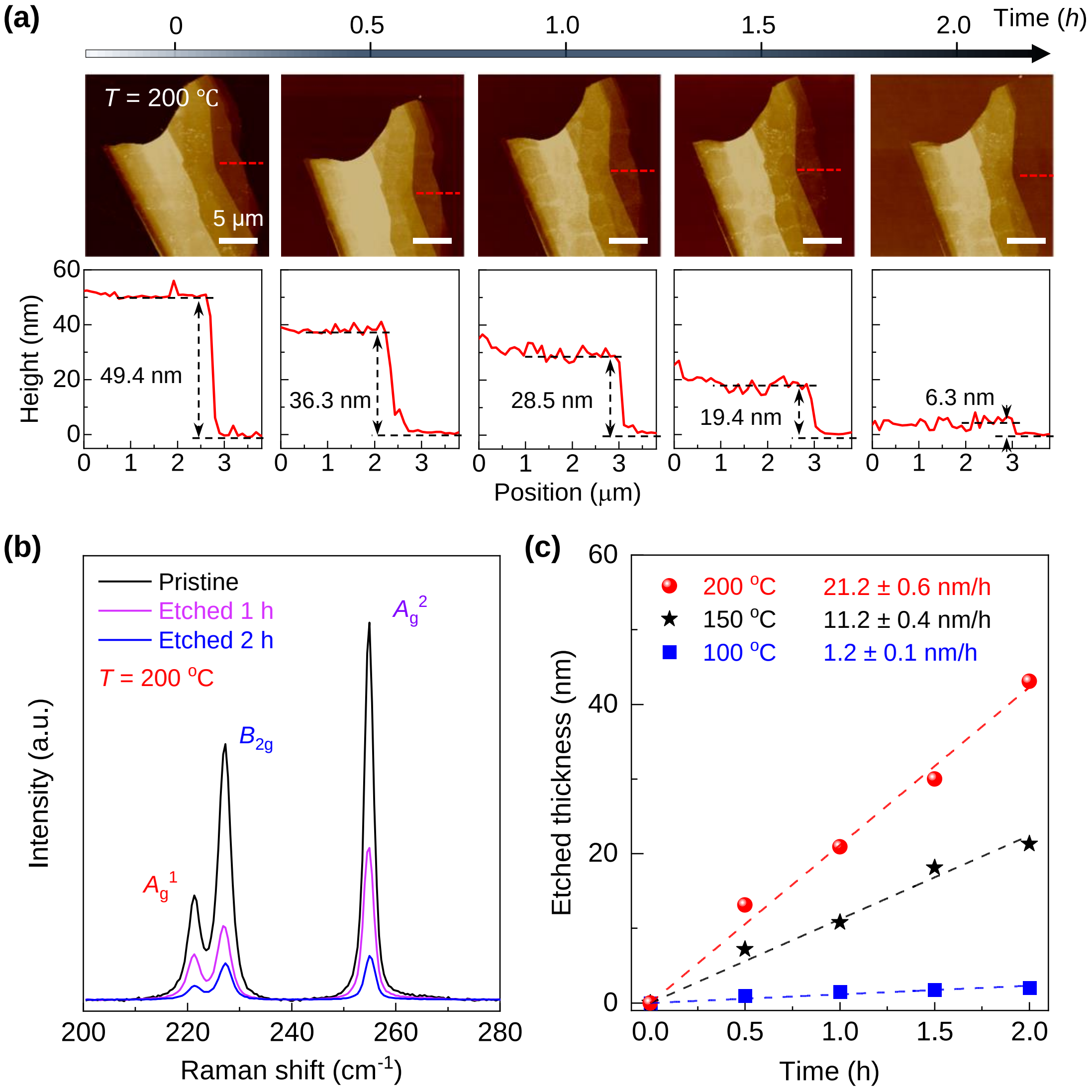

**Figure 4.** Temperature effect on the stability and thermal etching of b-As flake with controlled thickness. (a) AFM images of b-As flake under ambient atmosphere as a function of etching durations at 200 ºC (upper) and their corresponding height profiles at the same positions (lower). (b) Raman spectra of the pristine and etched b-As at 200 ºC. The peak intensity decreased significantly as the etching time lapsed. (c) Etched thickness of b-As flake as a function of etching duration at different temperatures of 100, 150 and 200 ºC. The linear etching rates are fitted as 1.2±0.1, 11.2±0.4, 21.2±0.6 nm/h at 100, 150 and 200 ºC, respectively.

Since temperature is a crucial variable in thermodynamics besides pressure, we also investigated the temperature effect on the structure and stability of b-As. It was found that thermal treatment can effectively modulate the chemical activity on the b-As surface, and hence accurately control the thickness of 2D b-As. As shown in Figure 4a, we first demonstrated the thickness evolution of a given b-As flake with temperature fixed at 200 ºC, while the annealing durations varies from 30, 60, 90 to 120 minutes, respectively. The flake thickness of 49.4 nm was effectively reduced to 6.3 nm in final, determined by atomic force microscopy (AFM) at the same position. It results in an etching rate of 21.1 nm/h. Raman spectroscopy was attempted to characterize the crystalline quality of b-As after thermal annealing. As the annealing time lapsed, it was found that three Raman vibration modes remained while their intensities dramatically decreased due to the reduced thickness at 200 ºC in Figure 4b. Importantly, no significant variation of the peak positions of these phonon modes arise, suggesting the negligible structural change during the entire thermal etching process. This phenomenon was observed in b-P as well.[57]

In order to reveal the etching mechanism, more control experiments were achieved. First, different temperatures were applied to perform the etching process, such as 100 and 150 °C. In kinetics, it is well known that lower temperature could obviously slow down the reaction rate. As shown in Figure 4c, we found that the linear slopes of etched thickness versus time are 1.2 ± 0.1 and 11.2 ± 0.4 nm/h at 100 and 150 °C, respectively, evidently lower than that at 200 °C as predicted. The etching rate increased by about 10 nm/h for every 50 °C interval in temperature. Therefore, it is a rational method to continuously and exactly control the thickness of b-As flake by optimizing the etching temperature and period. Also, it's best to keep the etching temperature below 300 °C, because when the etching temperature is above 300 °C, the etching rate is too fast, and thus it is difficult to preset the etching duration, as illustrated in Figure S3. The decomposition temperature of the b-As has been observed to be greater than 100 °C in ambient condition, which approaches the boiling point of arsenic acid. So, the etching mechanism maybe associated with the formation and thus evaporation of arsenic acid, consistent with that reported in b-P.[57] Second, we carried out thermal etching on b-As samples in an inert gas (argon) environment and found that the thickness of a given b-As flakes hardly changed after 2.5 h annealing at 300 °C. The results in Figure S4 indicate that oxygen or water in the environment play the pivotal role in the thermal etching of b-As. In addition, under the inert gas protection, b-As sublimated rapidly if the etching temperature is higher than 320 °C, as revealed in Figure S5.

In conclusion, we systematically investigated the structural metastability of the natural b-As and attempted to probe its thermodynamic origins, taking both hydrostatic pressure and thermal effect into account. Compared with b-P, it was found that b-As phase can only exist under an extremely narrow pressure range and irreversibly transform to g-As phase, suggesting the metastable nature of b-As in thermodynamics. This transition occurred with intralayer bond breakage and new interlayer bond formation, evidenced by TEM and DFT calculations. High temperature played the essential role to etch the b-As surface out and thus exactly control its thickness, while the crystalline structures were well retained. This study could boost many prospective applications of 2D materials in nanoelectronics and optoelectronics.

## EXPERIMENTAL METHODS

**High Pressure Measurements.** Hydrostatic pressure was applied by using a symmetric diamond anvil cell with 300 μm culet size. The pre-indented thickness of T301 gasket was around 50 μm to avoid the potential deformation and instability of the hole. Subsequently, a hole of 120 μm, as the sample chamber, was drilled by ultrafast laser in the center of gasket. Silicon oil was used as pressure-transmitting medium to offer the hydrostatic pressure condition. The pressure was calibrated by the photoluminescence of ruby balls, which are placed next to the b-As sample inside the pressure chamber.

**Temperature Effect on the Stability of B-As.** B-As flakes were mechanically exfoliated on polydimethylsiloxane using Scotch tape, and then the b-As flakes were transferred to $SiO_2$/Si substrate with 300 nm oxidation layer. To investigate temperature effect on the stability of b-As, the b-As sheets with known thickness were annealed at various temperatures (100, 150, 200, 300, and 320 ℃), and etching duration ranged from 0.5 to 2.0 h. Thermal etching experiments in air were carried out in muffle furnace with a nice temperature controller. In addition, chemical vapor deposition system was unitized to anneal the b-As in the inert Ar gas.

**Structural Characterizations of B-As Samples.** AFM (Dimension FastScan, Bruker) characterizations were performed in tapping mode. The thickness of b-As sheets was measured before and after thermal etching. Raman and PL spectra were acquired using a 633 nm laser as the excitation source and the iHR550 spectrometer. The laser power is low enough to avoid over-heating. For high-resolution TEM measurement, the exfoliated b-As flakes were transferred onto a holey carbon grid with polymer-free transfer technique. The detailed transfer procedure[58] can be found elsewhere. The characterizations were accomplished with FEI Titan TEM, and the accelerating voltage was 80 kV. Synchrotron XRD data were acquired at Beamline 12.2.2 of the Advanced Light Source (ALS) at Lawrence Berkeley National Laboratory. The beam energy was set as 18 keV (wavelength was around 0.6888 Å). The XRD signals were collected by a MAR345 image plate.

**DFT Calculations.** DFT calculations were performed by the Vienna Ab initio Simulation Package.[59, 60] The projector augmented wave[61] method was adopted to describe the core-valence interaction. The exchange-functional was treated by the generalized gradient approximation of Perdew-Burke-Ernzerhof functional.[62] The energy cut-off for the plane wave basis expansion was set to 500 eV and the convergence criterion of geometry relaxation was set to 0.03 eV/Å. The self-consistent calculations applied a convergence energy threshold of $10^{-5}$ eV. An 8×12×3 Monkhorst-Pack[63] *k*-point grid was used to sample the Brillouin zone of the *A*17 unit cell, and *k*-grids with similar density were used for other phases. The van der Waals interaction was included through the Grimmes's method (DFT-D2).[64] The minimum-energy pathway for phase transition was determined using the climbing-image nudged elastic band method.[65]

## ASSOCIATED CONTENT

Supporting Information is available free of charge online.

DFT calculation of the pressure-induced phase transitions of b-As; thermal treatment of b-As sheets under different temperatures; optical results of b-As flakes after thermal annealing; thermal treatment of b-As sheet under Ar atmosphere at 300 $^{o}$C (PDF)

## AUTHOR INFORMATION

### Corresponding Author

*E-mail: qlxia@csu.edu.cn, jkang@csrc.ac.cn, chyb0422@bit.edu.cn

### Author Contributions

Y.C. conceived the idea and led the project. G.D. performed the Raman characterizations and thermal annealing experiments. G.D., W.H. and Q.X. prepared the 2D samples. F.K. and B.C. carried out the high pressure XRD measurements. J.K. conducted the DFT calculations. Y.C. and G.D. analyzed the data and wrote the manuscript. All authors have given approval to the final version of the manuscript.

### Notes

The authors declare no competing financial interest.

## ACKNOWLEDGMENT

This work was financially supported by the National Natural Science Foundation of China (grant numbers 52072032, 12090031, 12274467, 12074029, 11991060, and U2230402) and the 173-JCJQ program (grant No. 2021-JCJQ-JJ-0159).

## REFERENCES

(1) Akinwande, D.; Brennan, C. J.; Bunch, J. S.; Egberts, P.; Felts, J. R.; Gao, H.; Huang, R.; Kim, J.-S.; Li, T.; Li, Y., A review on mechanics and mechanical properties of 2D materials—Graphene and beyond. *Extreme Mech. Lett.* **2017**, *13*, 42-77.

(2) Megra, Y. T.; Suk, J. W., Adhesion properties of 2D materials. *J. Phys. D: Appl. Phys.* **2019**, *52* (36), 364002.

(3) Naumis, G. G.; Barraza-Lopez, S.; Oliva-Leyva, M.; Terrones, H., Electronic and optical properties of strained graphene and other strained 2D materials: A review. *Rep. Prog. Phys.* **2017**, *80* (9), 096501.

(4) Li, X. L.; Han, W. P.; Wu, J. B.; Qiao, X. F.; Zhang, J.; Tan, P. H., Layer-number dependent optical properties of 2D materials and their application for thickness determination. *Adv. Funct. Mater.* **2017**, *27* (19), 1604468.

(5) Gupta, A.; Sakthivel, T.; Seal, S., Recent development in 2D materials beyond graphene. *Prog. Mater Sci.* **2015**, *73*, 44-126.

(6) Zhang, L.; Tang, Y.; Khan, A. R.; Hasan, M. M.; Wang, P.; Yan, H.; Yildirim, T.; Torres, J. F.; Neupane, G. P.; Zhang, Y.; Li, Q.; Lu, Y., 2D materials and heterostructures at extreme pressure. *Adv. Sci.* **2020**, *7* (24), 2002697.

(7) Novoselov, K. S.; Mishchenko, A.; Carvalho, A.; Castro Neto, A. H., 2D materials and van der Waals heterostructures. *Science* **2016**, *353* (6298), aac9439.

(8) Schaibley, J. R.; Yu, H.; Clark, G.; Rivera, P.; Ross, J. S.; Seyler, K. L.; Yao, W.; Xu, X., Valleytronics in 2D materials. *Nat. Rev. Mater.* **2016**, *1* (11), 1-15.

(9) Long, M.; Wang, P.; Fang, H.; Hu, W., Progress, challenges, and opportunities for 2D material based photodetectors. *Adv. Funct. Mater.* **2019**, *29* (19), 1803807.

(10) Zhang, G.; Zhang, Y.-W., Thermal properties of two-dimensional materials. *Chin. Phys. B* **2017**, *26* (3), 034401.

(11) Novoselov, K. S., Electric field effect in atomically thin carbon films. *Science* **2004**, *306* (5696), 666-669.

(12) Manzeli, S.; Ovchinnikov, D.; Pasquier, D.; Yazyev, O. V.; Kis, A., 2D transition metal dichalcogenides. *Nat. Rev. Mater.* **2017**, *2* (8), 17033.

(13) Ling, X.; Wang, H.; Huang, S.; Xia, F.; Dresselhaus, M. S., The renaissance of black phosphorus. *Proc. Nat. Acad. Sci.* **2015**, *112* (15), 4523-4530.

(14) Gogotsi, Y.; Anasori, B., The rise of MXenes. *ACS Nano* **2019**, *13* (8), 8491-8494.

(15) Akinwande, D.; Petrone, N.; Hone, J., Two-dimensional flexible nanoelectronics. *Nat. Commun.* **2014**, *5* (1), 5678.

(16) Liu, C.; Chen, H.; Wang, S.; Liu, Q.; Jiang, Y.-G.; Zhang, D. W.; Liu, M.; Zhou, P., Two-dimensional materials for next-generation computing technologies. *Nat. Nanotechnol.* **2020**, *15* (7), 545-557.

(17) Wang, X.; Sun, Y.; Liu, K., Chemical and structural stability of 2D layered materials. *2D Mater*. **2019**, *6* (4), 042001.

(18) Lanza, M.; Smets, Q.; Huyghebaert, C.; Li, L.-J., Yield, variability, reliability, and stability of two-dimensional materials based solid-state electronic devices. *Nat. Commun.* **2020**, *11* (1), 5689.

(19) Guo, Q.; Pospischil, A.; Bhuiyan, M.; Jiang, H.; Tian, H.; Farmer, D.; Deng, B.; Li, C.; Han, S.-J.; Wang, H.; Xia, Q.; Ma, T.-P.; Mueller, T.; Xia, F., Black phosphorus mid-infrared photodetectors with high gain. *Nano Lett.* **2016**, *16* (7), 4648-4655.

(20) Long, M.; Gao, A.; Wang, P.; Xia, H.; Ott, C.; Pan, C.; Fu, Y.; Liu, E.; Chen, X.; Lu, W.; Nilges, T.; Xu, J.; Wang, X.; Hu, W.; Miao, F., Room temperature high-detectivity mid-infrared photodetectors based on black arsenic phosphorus. *Sci. Adv.* **2017**, *3* (6), e1700589.

(21) Li, L.; Yu, Y.; Ye, G. J.; Ge, Q.; Ou, X.; Wu, H.; Feng, D.; Chen, X. H.; Zhang, Y., Black phosphorus field-effect transistors. *Nat. Nanotechnol.* **2014**, *9* (5), 372-7.

(22) Favron, A.; Gaufrès, E.; Fossard, F.; Phaneuf-L'Heureux, A.-L.; Tang, N. Y. W.; Lévesque, P. L.; Loiseau, A.; Leonelli, R.; Francoeur, S.; Martel, R., Photooxidation and quantum confinement effects in exfoliated black phosphorus. *Nat. Mater.* **2015**, *14* (8), 826-832.

(23) Island, J. O.; Steele, G. A.; Zant, H. S. J. v. d.; Castellanos-Gomez, A., Environmental instability of few-layer black phosphorus. *2D Mater.* **2015**, *2* (1), 011002.

(24) Abellán, G.; Wild, S.; Lloret, V.; Scheuschner, N.; Gillen, R.; Mundloch, U.; Maultzsch, J.; Varela, M.; Hauke, F.; Hirsch, A., Fundamental insights into the degradation and stabilization of thin layer black phosphorus. *J. Am. Chem. Soc.* **2017**, *139* (30), 10432-10440.

(25) Zhu, X.; Zhang, T.; Jiang, D.; Duan, H.; Sun, Z.; Zhang, M.; Jin, H.; Guan, R.; Liu, Y.; Chen, M.; Ji, H.; Du, P.; Yan, W.; Wei, S.; Lu, Y.; Yang, S., Stabilizing black phosphorus nanosheets via edge-selective bonding of sacrificial $C_{60}$ molecules. *Nat. Commun.* **2018**, *9* (1), 4177.

(26) Zhou, Q.; Chen, Q.; Tong, Y.; Wang, J., Light-induced ambient degradation of few-layer black phosphorus: mechanism and protection. *Angew. Chem. Int. Ed.* **2016**, *55* (38), 11437-11441.

(27) Gibertini, M.; Koperski, M.; Morpurgo, A. F.; Novoselov, K. S., Magnetic 2D materials and heterostructures. *Nat. Nanotechnol.* **2019**, *14* (5), 408-419.

(28) Zhang, T.; Grzeszczyk, M.; Li, J.; Yu, W.; Xu, H.; He, P.; Yang, L.; Qiu, Z.; Lin, H.; Yang, H.; Zeng, J.; Sun, T.; Li, Z.; Wu, J.; Lin, M.; Loh, K. P.; Su, C.; Novoselov, K. S.; Carvalho, A.; Koperski, M.; Lu, J., Degradation chemistry and kinetic stabilization of magnetic $CrI_3$. *J. Am. Chem. Soc*. **2022**, *144* (12), 5295-5303.

(29) Esters, M.; Hennig, R. G.; Johnson, D. C., Dynamic instabilities in strongly correlated $VSe_2$ monolayers and bilayers. *Phys. Rev. B* **2017**, *96* (23), 235147.

(30) Chen, Y.; Chen, C.; Kealhofer, R.; Liu, H.; Yuan, Z.; Jiang, L.; Suh, J.; Park, J.; Ko, C.; Choe, H. S.; Avila, J.; Zhong, M.; Wei, Z.; Li, J.; Li, S.; Gao, H.; Liu, Y.; Analytis, J.; Xia, Q.; Asensio, M. C.; Wu, J., Black arsenic: A layered semiconductor with extreme in-plane anisotropy. *Adv. Mater.* **2018**, *30* (30), e1800754.

(31) Zhong, M.; Meng, H.; Liu, S.; Yang, H.; Shen, W.; Hu, C.; Yang, J.; Ren, Z.; Li, B.; Liu, Y.; He, J.; Xia, Q.; Li, J.; Wei, Z., In-plane optical and electrical anisotropy of 2D black arsenic. *ACS Nano* **2021**, *15* (1), 1701-1709.

(32) Han, J.; Xie, J.; Zhang, Z.; Yang, D.; Si, M.; Xue, D., Negative Poisson's ratios in few-layer orthorhombic arsenic: First-principles calculations. *Appl. Phys. Express* **2015**, *8* (4), 041801.

(33) Du, Y.; Maassen, J.; Wu, W.; Luo, Z.; Xu, X.; Ye, P. D., Auxetic black phosphorus: A 2D material with negative Poisson's ratio. *Nano Lett.* **2016**, *16* (10), 6701-6708.

(34) Zhong, M.; Xia, Q.; Pan, L.; Liu, Y.; Chen, Y.; Deng, H.-X.; Li, J.; Wei, Z., Thickness-dependent carrier transport characteristics of a new 2D elemental semiconductor: Black arsenic. *Adv. Funct. Mater.* **2018**, *28* (43), 1802581.

(35) Degtyareva, O.; McMahon, M. I.; Nelmes, R. J., High-pressure structural studies of group-15 elements. *High Pressure Res.* **2004**, *24* (3), 319-356.

(36) Needs, R. J.; Martin, R. M.; Nielsen, O. H., Total-energy calculations of the structural properties of the group-V element arsenic. *Phys. Rev. B* **1986**, *33* (6), 3778-3784.

(37) Li, X.; Sun, J.; Shahi, P.; Gao, M.; MacDonald, A.; Uwatoko, Y.; Xiang, T.; Goodenough, J.; Cheng, J. G.; Zhou, J. S., Pressure-induced phase transitions and superconductivity in a black phosphorus single crystal. *Proc. Nat. Acad. Sci.* **2018**, *115*, 201810726.

(38) Beister, H. J.; Strossner, K.; Syassen, K., Rhombohedral to simple-cubic phase transition in arsenic under pressure. *Phys. Rev. B* **1990**, *41* (9), 5535-5543.

(39) Gao, C.; Li, R.; Zhong, M.; Wang, R.; Wang, M.; Lin, C.; Huang, L.; Cheng, Y.; Huang, W., Stability and phase transition of metastable black arsenic under high pressure. *J. Phys. Chem. Lett.* **2020**, *11* (1), 93-98.

(40) Nan, H. Y.; Ni, Z. H.; Wang, J.; Zafar, Z.; Shi, Z. X.; Wang, Y. Y., The thermal stability of graphene in air investigated by Raman spectroscopy. *J. Raman Spectrosc.* **2013**, *44* (7), 1018-1021.

(41) Liu, X.; Wood, J. D.; Chen, K.-S.; Cho, E.; Hersam, M. C., In situ thermal decomposition of exfoliated two-dimensional black phosphorus. *J. Phys. Chem. Lett.* **2015**, *6* (5), 773-778.

(42) Wood, J. D.; Wells, S. A.; Jariwala, D.; Chen, K.-S.; Cho, E.; Sangwan, V. K.; Liu, X.; Lauhon, L. J.; Marks, T. J.; Hersam, M. C., Effective passivation of exfoliated black phosphorus transistors against ambient degradation. *Nano Lett.* **2014**, *14* (12), 6964-6970.

(43) Feng, X.; Kulish, V. V.; Wu, P.; Liu, X.; Ang, K.-W., Anomalously enhanced thermal stability of phosphorene via metal adatom doping: An experimental and first-principles study. *Nano Res.* **2016**, *9* (9), 2687-2695.

(44) Osters, O.; Nilges, T.; Bachhuber, F.; Pielnhofer, F.; Weihrich, R.; Schoneich, M.; Schmidt, P., Synthesis and identification of metastable compounds: black arsenic--science or fiction? *Angew. Chem. Int. Ed. Engl.* **2012**, *51* (12), 2994-7.

(45) Norman, N. C., Chemistry of arsenic, antimony and bismuth. Springer Science & Business Media, Germany, 1997.

(46) Yoshiasa, A.; Tokuda, M.; Misawa, M.; Shimojo, F.; Momma, K.; Miyawaki, R.; Matsubara, S.; Nakatsuka, A.; Sugiyama, K., Natural arsenic with a unique order structure: potential for new quantum materials. *Sci. Rep.* **2019**, *9* (1), 6275.

(47) Liu, Y.; Wang, T.; Robertson, J.; Luo, J.; Guo, Y.; Liu, D., Band structure, band offsets, and intrinsic defect properties of few-layer arsenic and antimony. *J. Phys. Chem. C* **2020**, *124* (13), 7441-7448.

(48) Sturala, J.; Sofer, Z.; Pumera, M., Coordination chemistry of 2D and layered gray arsenic: photochemical functionalization with chromium hexacarbonyl. *NPG Asia Mater.* **2019**, *11* (1), 42.

(49) Seidl, M.; Balázs, G.; Scheer, M., The chemistry of yellow arsenic. *Chem. Rev.* **2019**, *119* (14), 8406-8434.

(50) Kamal, C.; Ezawa, M., Arsenene: Two-dimensional buckled and puckered honeycomb arsenic systems. *Phys. Rev. B* **2015**, *91* (8), 085423.

(51) Zhu, Y.; Zheng, W.; Wang, W.; Zhu, S.; Cheng, L.; Li, L.; Lin, Z.; Ding, Y.; Jin, M.; Huang, F., Raman tensor of layered black arsenic. *J. Raman Spectrosc.* **2020**, *51* (8), 1324-1330.

(52) Saboori, S.; Deng, Z.; Li, Z.; Wang, W.; She, J., β-As Monolayer: Vibrational properties and Raman spectra. *ACS Omega* **2019**, *4* (6), 10171-10175.

(53) Vishnoi, P.; Mazumder, M.; Pati, S. K.; R. Rao, C. N., Arsenene nanosheets and nanodots. *New J. Chem.* **2018**, *42* (17), 14091-14095.

(54) Xiao, G.; Cao, Y.; Qi, G.; Wang, L.; Zeng, Q.; Liu, C.; Ma, Z.; Wang, K.; Yang, X.; Sui, Y.; Zheng, W.; Zou, B., Compressed few-layer black phosphorus nanosheets from semiconducting to metallic transition with the highest symmetry. *Nanoscale* **2017**, *9* (30), 10741-10749.

(55) Shih, C.; Hu, S.; Chen, M.; Huang, S., Randomness-induced evolution of the first-order to the second-order phase transition in two-dimensional six-state potts model. *Life Sci. J.* **2009**, *6* (2), 29-32.

(56) Shang, S.; Wang, Y.; Zhang, H.; Liu, Z.-K., Lattice dynamics and anomalous bonding in rhombohedral As: First-principles supercell method. *Phys. Rev. B* **2007**, *76* (5), 052301.

(57) Jeong, M.-H.; Kwak, D.-H.; Ra, H.-S.; Lee, A. Y.; Lee, J.-S., Realizing long-term stability and thickness control of black phosphorus by ambient thermal treatment. *ACS Appl. Mater. Interfaces* **2018**, *10* (22), 19069-19075.

(58) Regan, W.; Alem, N.; Alemán, B.; Geng, B.; Girit, Ç.; Maserati, L.; Wang, F.; Crommie, M.; Zettl, A., A direct transfer of layer-area graphene. *Appl. Phys. Lett.* **2010**, *96* (11), 113102.

(59) Kresse, G.; Hafner, J., Ab initio molecular dynamics for liquid metals. *Phys. Rev. B* **1993**, *47* (1), 558-561.

(60) Kresse, G.; Furthmüller, J., Efficient iterative schemes for ab initio total-energy calculations using a plane-wave basis set. *Phys. Rev. B* **1996**, *54* (16), 11169-11186.

(61) Kresse, G.; Joubert, D., From ultrasoft pseudopotentials to the projector augmented-wave method. *Phys. Rev. B* **1999**, *59* (3), 1758-1775.

(62) Perdew, J. P.; Burke, K.; Ernzerhof, M., Generalized gradient approximation made simple. *Phys. Rev. Lett.* **1996**, *77* (18), 3865-3868.

(63) Monkhorst, H. J.; Pack, J. D., Special points for Brillouin-zone integrations. *Phys. Rev. B* **1976**, *13* (12), 5188-5192.

(64) Grimme, S.; Antony, J.; Ehrlich, S.; Krieg, H., A consistent and accurate ab initio parametrization of density functional dispersion correction (DFT-D) for the 94 elements H-Pu. *J. Chem. Phys.* **2010**, *132* (15), 154104.

(65) Henkelman, G.; Uberuaga, B. P.; Jónsson, H., A climbing image nudged elastic band method for finding saddle points and minimum energy paths. *J. Chem. Phys.* **2000**, *113* (22), 9901-9904.